\documentclass[acmsmall,screen,authorversion,nonacm]{acmart}

\usepackage{amsmath,amsfonts}
\usepackage{algorithm}
\usepackage{algorithmic}
\usepackage[caption=false,font={small},labelfont=sf,textfont=sf]{subfig}
\usepackage[font={small},labelfont={bf}]{caption}
\usepackage{subcaption}
\usepackage{multirow}
\usepackage{array}
\usepackage{enumitem}
\usepackage{fancyhdr}
\AtBeginDocument{%
    \addtolength{\footskip}{2.0\baselineskip}%
    \fancyfoot[L]{\textit{\textbf{Preprint --- do not distribute.}}}%
}

\usepackage{pifont} 
\AtBeginDocument{%
  }

\setcopyright{none}
    
\copyrightyear{2025}
\acmYear{2025}
\setcopyright{rightsretained}
\acmConference[ASSETS '25]{The 27th International ACM SIGACCESS Conference on Computers and Accessibility}{October 26--29, 2025}{Denver, CO, USA}
\acmBooktitle{The 27th International ACM SIGACCESS Conference on Computers and Accessibility (ASSETS '25), October 26--29, 2025, Denver, CO, USA}
\acmDOI{10.1145/3663547.3759726}
\acmISBN{979-8-4007-0676-9/2025/10}

\begin{document}

\title{NaviSense: A Multimodal Assistive Mobile application for Object Retrieval by Persons with Visual Impairment}


\author{Ajay Narayanan Sridhar}
\affiliation{%
  \institution{The Pennsylvania State University}
  \city{University Park, PA}
  \country{USA}}
\email{afs6372@psu.edu}

\author{Fuli Qiao}
\affiliation{%
  \institution{The Pennsylvania State University}
  \city{University Park, PA}
  \country{USA}}
\email{fvq5015@psu.edu}

\author{Nelson Daniel Troncoso Aldas}
\affiliation{%
  \institution{Independent Researcher}
  \city{Stanford, CA}
  \country{USA}}
  \email{nd.cse@proton.me}

\author{Yanpei Shi}
\affiliation{%
  \institution{University of Southern California}
  \city{Los Angeles, CA}
  \country{USA}}
\email{yanpeish@usc.edu}

\author{Mehrdad Mahdavi}
\affiliation{%
  \institution{The Pennsylvania State University}
  \city{University Park, PA}
  \country{USA}}
\email{mzm616@psu.edu}

\author{Laurent Itti}
\affiliation{%
  \institution{University of Southern California}
  \city{Los Angeles, CA}
  \country{USA}}
\email{itti@usc.edu}

\author{Vijaykrishnan Narayanan}
\affiliation{%
  \institution{The Pennsylvania State University}
  \city{University Park, PA}
  \country{USA}}
\email{vxn9@psu.edu}

\renewcommand{\shortauthors}{Ajay Narayanan Sridhar, Fuli Qiao, et al.}

\begin{abstract}
People with visual impairments often face significant challenges in locating and retrieving objects in their surroundings. Existing assistive technologies present a trade-off: systems that offer precise guidance typically require pre-scanning or support only fixed object categories, while those with open-world object recognition lack spatial feedback for reaching the object. To address this gap, we introduce \textit{NaviSense}, a mobile assistive system that combines conversational AI, vision-language models, augmented reality (AR), and LiDAR to support open-world object detection with real-time audio-haptic guidance. Users specify objects via natural language and receive continuous spatial feedback to navigate toward the target without needing prior setup. Designed with insights from a formative study and evaluated with 12 blind and low-vision participants, NaviSense significantly reduced object retrieval time and was preferred over existing tools, demonstrating the value of integrating open-world perception with precise, accessible guidance.

\end{abstract}



\begin{teaserfigure}
  \centering
  \includegraphics[width=\linewidth]{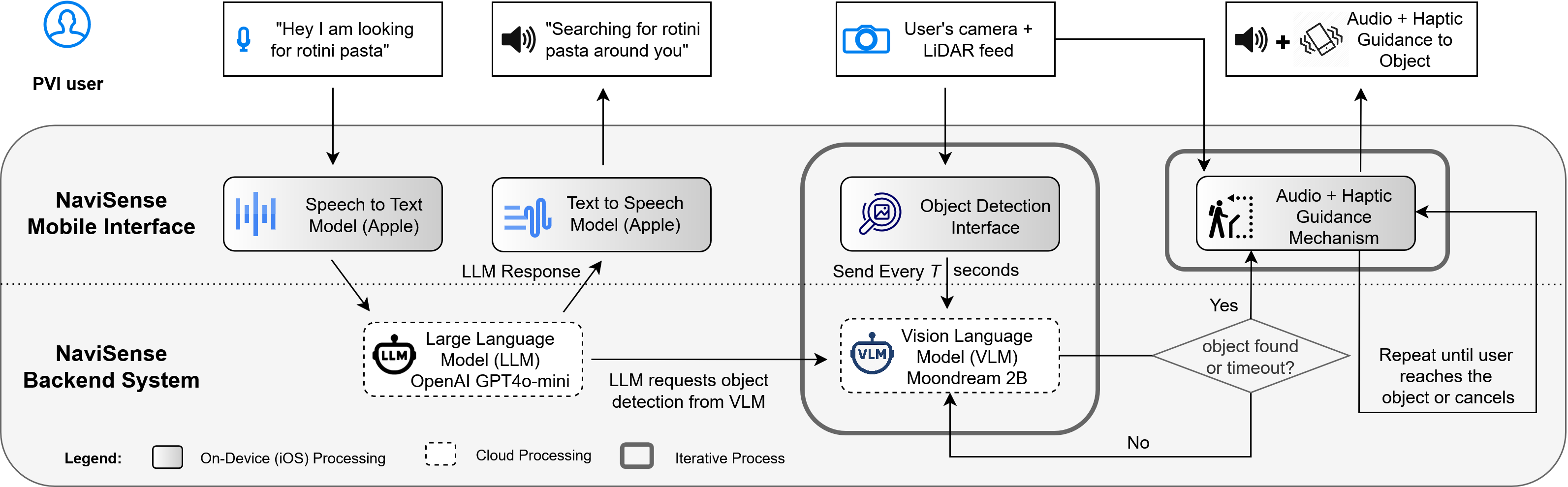} 
  \caption{System architecture of NaviSense.   Gradient-filled boxes indicate on-device processing components (e.g., speech recognition and synthesis); dashed-border boxes denote cloud-based processing (LLM and VLM); thick-border boxes represent iterative operations. The system uses speech input, real-time camera and LiDAR feeds, and provides multimodal feedback (audio and haptic) to guide blind or visually impaired users to the requested object.}
  \Description{A system diagram for NaviSense, a navigation tool for blind and visually impaired users. The diagram is split into three rows: the user interaction (on the top), mobile interface (in the middle) and the backend system (at the bottom). A user speaks a query, such as “I am looking for rotini pasta,” which is converted to text using an on-device Apple speech-to-text model. This text is sent to a cloud-based large language model (OpenAI GPT-4o-mini), which determines the object to locate and queries a cloud-based vision-language model (Moondream 2B). Meanwhile, the user’s camera and LiDAR feed are processed locally using an object detection interface, sending data to the vision-language model every few seconds. When the object is identified, the system provides audio and haptic guidance to the user using on-device processing. Guidance continues until the user reaches the object or cancels. Gradient-filled boxes indicate on-device (iOS) processing, dashed-border boxes represent cloud-based processing, and thick-border boxes represent iterative operations.}
  \label{fig:navisense-architecture}
\end{teaserfigure}

\maketitle

\section{Introduction}

People with visual impairments (PVI) often struggle to independently locate and retrieve everyday objects~\cite{el2023survey}, impacting their autonomy and frequently necessitating external assistance~\cite{townley2009understanding, mcclimens2014people, lubin2012role}. Tasks such as identifying groceries or finding personal belongings typically require either direct human help or cumbersome pre-arrangements~\cite{benton1993visuoperceptual}. Remote-sighted assistance platforms such as Be My Eyes~\cite{bemyeyes}, VizWiz::LocateIt~\cite{bigham2010vizwiz}, and Aira~\cite{aira} provide real-time human support, allowing users to receive visual feedback via volunteers or paid assistants. While effective, these systems introduce issues related to privacy, availability, and financial cost, thereby restricting spontaneous and independent use.

To overcome these limitations, automated computer-vision-based systems~\cite{aipoly, ProgramAlly, taptapsee} have been developed. However, these systems often face a critical trade-off. On one hand, tools like AIGuide~\cite{troncoso2020aiguide}, SeeingAI~\cite{seeingai}, and Recog~\cite{ahmetovic2020recog} provide precise guidance but only for a limited set of pre-defined or pre-scanned objects. On the other hand, newer systems with open-world object detection, such as Be My AI~\cite{bemyeyes_introducing_be_my_ai2023}, WorldScribe~\cite{chang2024worldscribe}, OpenAI Advanced Voice Mode~\cite{openai_advancedvoice2024}, and Gemini Live~\cite{google_gemini_live2025}, offer broader recognition capabilities but mainly provide descriptive feedback without guiding users to the object. As a result, current solutions often require users to choose between flexibility and navigational utility.

Recent advancements in artificial intelligence, including large language models (LLMs)\cite{yu2023mm,yang2024viassist}, vision-language models (VLMs)\cite{ahmetovic2020recog}, augmented reality (AR)\cite{xie2022helping}, and mobile LiDAR sensors\cite{di2021mobile,cheng2018mobile}, have created opportunities to address these gaps effectively~\cite{stone2023open, cao2023mobile}. Leveraging these technological innovations and insights from prior works, we conducted a formative study with 12 visually impaired participants to deeply understand their needs and preferences regarding assistive object retrieval systems.

Building upon these insights, we developed \textit{NaviSense}, a conversational AI-based assistive mobile application designed to support independent, real-time object retrieval. NaviSense employs an LLM-driven conversational interface combined with VLM-based open-world object identification. It utilizes AR and LiDAR technologies to precisely localize objects within the user's environment. Upon detection, NaviSense provides continuous, multimodal guidance through synchronized audio instructions and haptic feedback, enabling users to confidently and accurately reach and interact with target objects. We validated NaviSense via a comparative user study with 12 participants completing structured object retrieval tasks using NaviSense, Be My AI~\cite{bemyeyes_introducing_be_my_ai2023}, and Ray-Ban Meta Smart Glasses~\cite{raybanmeta}. Results demonstrated significantly higher user satisfaction and task success rates with NaviSense, highlighting the importance of real-time multimodal feedback and conversational interactions as critical facilitators of user independence. 
\section{Related Works}

\noindent\textbf{Object Finding approaches for PVI users:}
Traditional mobility aids like white canes and guide dogs effectively assist with obstacle detection and navigation, typically without supporting locating and grasping objects. To address this limitation, remote-sighted assistance platforms, including Be My Eyes~\cite{bemyeyes}, VizWiz::LocateIt~\cite{bigham2010vizwiz}, and Aira~\cite{aira}, provide real-time human visual assistance but are limited by availability, cost, and privacy concerns. Smartphone applications like Be My AI~\cite{bemyeyes_introducing_be_my_ai2023}, WorldScribe~\cite{chang2024worldscribe}, and others~\cite{taptapsee, openai_advancedvoice2024, google_gemini_live2025, seeingai, lee2022imageexplorer} offer automated object and scene recognition. However, these solutions generally lack precise guidance for physically interacting with identified objects. Specialized wearables like Take My Hand~\cite{rahman2023take}, WanderGuide~\cite{kuribayashi2025wanderguide}, Ray-Ban Meta Glasses~\cite{raybanmeta}, and others~\cite{liu2025objectfinderopenvocabularyassistiveinteractive, mathis2025lifeinsight, thakoor2015system} enable innovative object interaction capabilities but are often costly or cumbersome. Recent works like ThirdEye~\cite{third-eye} and AIGuide~\cite{troncoso2020aiguide} target the ``last-metre'' problem~\cite{manduchi2014last} with AR and haptics for hand-level guidance but rely on pre-scanned objects, limiting generalization. 
Personal object finders such as Recog~\cite{ahmetovic2020recog} and SeeingAI~\cite{seeingai} rely on object pre-registration and struggle in dynamic environments. Our NaviSense builds on these efforts with an open-world object detection pipeline that identifies and guides users to objects without requiring pre-scanning.

\noindent\textbf{Multimodal Non-Visual Directional Guidance Interfaces:}
Multimodal interfaces support navigation and object retrieval for people with visual impairments (PVI). Auditory systems deliver spatial cues through verbal instructions~\cite{duh2020v, white2010easysnap, thakoor2015system} and/or non-verbal audio signals~\cite{hu2022stereopilot, dramas2008designing, ahmetovic2020recog, ahmetovic2016navcog}, while may falter in noisy settings. Haptic feedback systems, using vibrotactile~\cite{scheggi2014remote, bhatlawande2013way, de2008bringing} or skin-stretch cues~\cite{kayhan2022wearable}, provide tactile guidance through wearable devices like gloves~\cite{third-eye, zelek2003haptic}, finger-worn devices~\cite{horvath2014fingersight, stearns2016evaluating} and handheld devices~\cite{pielot2011tactile}, but often require extensive calibration and specialized hardware. To leverage the strengths of both auditory and haptic modalities, multimodal systems such as AIGuide~\cite{troncoso2020aiguide}, Find My Things~\cite{wen2024find}, and NaviGPT~\cite{zhang2025enhancing} integrate audio instructions and haptic cues for effective guidance, and are shown to be effective compared to single modality~\cite{kim2016assisting, lee2017reaching, lee2022aiguide}. Inspired by these, NaviSense employs a multimodal feedback approach combining verbal prompts, auditory tones, and vibration feedback, adapting dynamically to real-time user behavior and environmental contexts to facilitate precise object interaction.

\section{Design Procedures}

\subsection{User Study}
To inform NaviSense's design, we interviewed 12 visually impaired participants (9 blind, 3 low vision; 8 females, 4 males; aged 39–73, $M=56.5$, $SD=9.4$), recruited via local blindness organizations and compensated \$30/hour. Zoom interviews (45 mins-1 hour) explored participants’ current strategies, frustrations, and expectations for daily indoor object finding. We used open coding to extract user needs and system requirements, alongside insights from prior work.

\subsection{Design Considerations}
\noindent\textbf{\textit{D1: Open-world Object Detection.}} Participants noted failures of existing systems in recognizing everyday items, particularly items with similar packaging. NaviSense addresses this via an open-vocabulary object detection model powered by a vision-language~\cite{liu2023visual} pipeline, allowing flexible detection based on object description. \\
\noindent\textbf{\textit{D2: Precise Guidance and Feedback.}}  Users often lacked step-by-step assistance to actually locate or touch objects, even when they had been identified. Thus, we implemented a multi-modal guidance loop using real-time audio and haptic cues, leveraging LiDAR-based depth sensing on iPhones to guide users' hands toward the object’s position. \\\noindent\textbf{\textit{D3: Conversational Interface.}}  Participants strongly preferred natural voice interaction over screen-based or typing interfaces. Hence, we integrated speech-based input and output alongside a language model (GPT-4o-mini), enabling fluid, hands-free interactions for querying and navigating the system.

\section{Implementation}

NaviSense comprises an iOS application that uses ARKit~\cite{apple_arkit_2025}, camera, microphone, and LiDAR sensors, with a cloud backend hosting an LLM for voice interaction and a VLM for object detection (see Fig.~\ref{fig:navisense-architecture}). Development and testing were conducted using an iPhone 16 Pro and an Apple M2 MacBook Pro (16 GB RAM). The app requires permissions for the internet, camera, microphone, and speech recognition. Its core components are:
\begin{enumerate}
    \item \textbf{~Conversational Voice Interface: } The user initiates interactions verbally (e.g., "Find my coffee cup"). Apple's on-device Speech-to-Text converts audio to text, sending it to OpenAI GPT-4o-mini~\cite{openai_gpt4omini} to interpret and clarify user intent conversationally.
    
    \item \textbf{~3D Object Detection: }
    Upon receiving the user's query, a video frame is sent to Moondream 2B~\cite{moondream_vlm}, an open-world VLM. It detects the requested object visually without pre-defined categories. The identified object's 2D location is sent back to the device, converted into a 3D spatial point using ARKit and LiDAR depth data. This spatial position is continuously tracked and updated locally.

    \item \textbf{~Multimodal Guidance Feedback (Audio-Haptic):} NaviSense translates spatial data into actionable feedback:
        \begin{itemize}[topsep=1pt,itemsep=1pt,parsep=0pt,partopsep=0pt,left=10pt]
        \item \textbf{Audio Feedback}: Directional voice prompts guide the user toward objects.
        \item \textbf{Haptic Feedback}: Distance-based vibrations signal proximity—slow pulses at a distance, faster pulses when approaching, and rapid pulses upon nearing the object.
        \end{itemize}
    
\end{enumerate}

\section{Evaluation}

\subsection{User Study}

We conducted a user study with 12 PVI participants (P1–P12) to assess \textbf{NaviSense} in helping users locate and retrieve everyday objects. Each 2-hour session included retrieval tasks using three systems: NaviSense, Be My AI, and Ray-Ban Meta Smart Glasses, followed by a brief semi-structured interview. \\
\textbf{Study Procedure.} We adopted a similar application testing setup as that of AIGuide~\cite{troncoso2020aiguide}. Participants retrieved three target objects (3 trials per object): A2 Milk Carton~\cite{a2_milk}, party cups~\cite{great_value_products}, and rotini pasta box~\cite{great_value_products}, from a four-tier shelf stocked with eight household items, chosen to vary in visual and tactile properties.
Tasks were performed using each system (18 trials in total), with randomized object positions and counterbalanced trial order. For each trial, we recorded five metrics: \textit{search time} (to locate the object), \textit{guidance time} (to reach and pick it up), \textit{total time} (combined duration), \textit{accuracy} (successful retrievals), and \textit{undesired object touches} (mistaken interactions). After completing the tasks, participants gave feedback on system usability, guidance clarity and general experience.

\begin{table}[ht]
\small
\centering
\caption{Performance metrics across methods. Lower values indicate better performance for all metrics except accuracy.\vspace{-1em}}
\label{tab:performance_summary}
\begin{tabular}{lccc}
\toprule
\textbf{Metric} & \textbf{Be My AI} & \textbf{Meta Glasses} & \textbf{NaviSense (ours)} \\
\midrule
Search Time (s) & 48.23 $\pm$ 19.64 & 21.45 $\pm$ 14.58 & \textbf{15.86} $\mathbf{\pm}$ \textbf{5.65} \\
Guidance Time (s) & 15.97 $\pm$ 12.85 & 19.35 $\pm$ 14.23 & \textbf{15.89} $\mathbf{\pm}$ \textbf{6.04} \\
Total Time (s) & 64.19 $\pm$ 25.01 & 40.80 $\pm$ 20.83 & \textbf{31.75} \textbf{$\pm$} \textbf{8.11} \\
Undesired Objects & 2.65 $\pm$ 2.07 & 4.37 $\pm$ 2.69 & \textbf{0.52} $\mathbf{\pm}$ \textbf{0.85 }\\
Accuracy (\%) & 85.71 & 55.88 & \textbf{95.37} \\
\bottomrule
\end{tabular}
\end{table}

\subsection{Quantitative Analysis}

Tab.~\ref{tab:performance_summary} summarizes performance metrics across participants. Repeated-measures ANOVA showed significant effects on search time (\(F = 77.59, p < 0.001\)) and total time (\(F = 34.46, p < 0.001\)), but not guidance time (\(F = 0.86, p = 0.44\)). Paired t-tests with Bonferroni correction (\(\alpha = 0.017\)) confirmed NaviSense significantly reduced search time vs. Be My AI (\(p < 10^{-8}\)); the difference vs. Meta Glasses (\(p = 0.046\)) was not significant. For total time, NaviSense outperformed both Meta Glasses (\(p = 0.0095\)) and Be My AI (\(p < 10^{-5}\)). A Friedman test found significant differences in undesired touches (\(\chi^2 = 18.5, p < 0.001\)); post-hoc Wilcoxon tests showed NaviSense had fewer errors than Meta Glasses (\(p = 0.0005\)) and Be My AI (\(p = 0.001\)). Accuracy was highest for NaviSense at \textbf{95.4\%} (95\% CI   [0.896, 0.980]), followed by Be My AI (\textbf{85.7\%}, CI [0.778, 0.911]) and Meta Glasses (\textbf{55.9\%}, CI [0.462, 0.651]). A two-way ANOVA found significant differences in total retrieval times across methods (\(F = 33.77, p < 0.001\)), but no effect of object type or method-object interaction.

\subsection{Qualitative Analysis}
\label{sec:qualitative_analysis}

After completing object retrieval tasks, participants completed semi-structured interviews that included 1-5 Likert-scale questions and open-ended prompts to evaluate usability, feedback clarity, perceived accuracy, and system preferences.\\
\textbf{Rating Questions Analysis. }
Participants expressed strong satisfaction with NaviSense’s design and functionality, especially its audio-haptic feedback ($4.67\pm0.49$) and ease of use ($4.42\pm0.90$). Perceived accuracy was also rated highly ($4.25\pm1.14$), though a few participants noted occasional confusion when feedback was delivered too quickly. Participants overwhelmingly preferred NaviSense over the other tools. It received the highest rating ($4.63\pm0.55$), significantly outperforming Be My AI ($3.81\pm1.00$) and Meta Glasses ($1.90\pm0.83$), with a significant effect of assistive method confirmed by a Friedman test (\(\chi^2(2) = 20.83, p < 0.001\)). Post-hoc Wilcoxon tests with Bonferroni correction \((\alpha = 0.017)\) showed NaviSense was rated higher than both Be My AI ($p = 0.011$) and Meta Glasses ($p = 0.001$).\\
\noindent\textbf{Open-Ended Questions Analysis. }
Thematic analysis of open-ended responses revealed three key themes: \textit{(1) Effective Multimodal Feedback:} Participants consistently praised the spatial clarity of NaviSense’s audio and haptic cues. As P7 remarked, \textit{“The beeping sound... really helps you zero in on object locations.”}. \textit{(2) Desire for Integration and Customization:} Participants expressed a desire for a single system combining features from all three tools: hands-free input, directional guidance, and concise object descriptions. P6 noted, \textit{“I wish there was just one device that did it all.”} Others (P7, P8, P9, P11) suggested customizations like gesture-based cancellation. \textit{(3) Expanded Everyday Use:} Participants imagined future use cases such as identifying freezer contents or locating luggage at the airport. P8 shared, \textit{“It could tell me what I pull out of my freezer… find my luggage at the airport.”}

\noindent Overall, participants viewed NaviSense positively, highlighting its intuitive feedback while suggesting enhancements for flexibility and real-world applicability.

\subsection{Evaluation on Everyday Object Categories}
\label{sec:ecovalidity}

To assess NaviSense’s generalizability beyond the initial object set, we conducted a supplementary evaluation using 25 commonly searched household items identified through formative interviews with PVI users. These objects were grouped into five categories: groceries, personal items, medicines, household products, and clothing. For each category, we recorded cluttered 8-second video scenes and sampled 200 frames overall. NaviSense's detection pipeline achieved 95\% accuracy (95\% CI = [0.911, 0.974]), correctly identifying or ruling out target objects in most frames. Errors included two false positives (e.g., mistaking penne for rotini pasta) and eight false negatives, mainly due to occlusion (e.g., dishwasher pods). These results demonstrate NaviSense’s robustness in realistic, visually complex settings, supporting its ecological validity, though further testing in more varied conditions is needed.

\section{Discussion and Conclusion}

Our study highlights key design trade-offs among three object recognition tools for blind and visually impaired users. Meta Glasses offered a hands-free form factor but lacked response consistency and natural language flexibility. Be My AI provided useful scene-level descriptions but lacked spatial specificity, often leading to information overload. In contrast, NaviSense enabled more independent and efficient object retrieval by combining open-world object detection with real-time multimodal guidance. Clear onboarding and spatial framing were keys, as participants quickly adapted once they understood the guidance based on the phone’s rear camera.

Despite promising results, our study has limitations. The participant pool was small ($N=12$) and demographically narrow, limiting generalizability. Tasks were restricted to a subset of household items, and detection challenges remain for occluded or visually similar objects. NaviSense requires line-of-sight interaction and steady phone handling, which may be physically demanding, and relies on cloud-based models, raising privacy and connectivity concerns. Future work should explore on-device implementations, hands-free interaction, customizable feedback, and broader assistive tasks like label reading, scene interpretation, and basic task planning.

In summary, we introduced NaviSense, a conversational, AR-based mobile assistive system that enables blind and visually impaired users to locate objects using natural language and receive real-time audio-haptic guidance. Informed by user feedback and validated in a comparative user study, NaviSense significantly improved object search performance and user satisfaction. These findings demonstrate the potential of integrating open-world perception with precise, accessible guidance to support everyday independence for people with visual impairments.

\begin{acks}
This material is based upon work supported by the National Science Foundation (NSF) under Grant Number 2318101. Any opinions, findings, and conclusions or recommendations expressed in this material are those of the author(s) and do not necessarily reflect the views of the National Science Foundation. We sincerely thank our participants with visual impairments for their valuable contributions to the user study.
\end{acks}

\bibliographystyle{ACM-Reference-Format}
\bibliography{sample-base}

\appendix

\section{Appendix: Real-World Scenarios Illustrating NaviSense’s Differentiation}

We highlight how NaviSense differs from existing solutions through two common, real-world scenarios that illustrate its unique combination of open-world object detection, real-time hand guidance, and conversational interaction.

\textbf{Scenario 1: Finding Dropped Keys.}  
Imagine Brian, a blind individual, who hears the faint sound of something falling while getting ready to leave home. He suspects it might be his keys, but having missed the direction of the sound, he has no idea where they landed. Without assistive tools, Brian might crouch and sweep the floor with his hands, relying on touch to locate them. Alternatively, he could call a family member via FaceTime or connect with a volunteer on Be My Eyes~\cite{bemyeyes} or Aira~\cite{aira}, depending on someone else's availability and sight.

AI-powered vision apps like Be My AI~\cite{bemyeyes}, WorldScribe~\cite{chang2024worldscribe}, or similar tools~\cite{openai_advancedvoice2024, google_gemini_live2025, ProgramAlly, lee2022imageexplorer} might announce that they "see keys" somewhere in the frame. However, Brian would still have to guess the precise location and feel around to find them. These systems provide descriptive feedback but lack step-by-step guidance or spatial precision.

Other tools like Find My Things~\cite{ahmetovic2020recog} or AIGuide~\cite{lee2022aiguide} do support fine-grained hand guidance through audio and haptic feedback. However, they require the object to be pre-scanned and labeled in advance; a limitation when the object, like Brian’s keys, was not previously registered.

With \textbf{NaviSense}, Brian simply opens the app and says, “Find my keys,” or adds helpful detail, such as, “Find my keys with the red keychain.” If needed, the app asks clarifying questions to improve accuracy. As he scans the environment, NaviSense uses open-world object detection (without any prior training) to locate the keys. Once found, the app delivers real-time, multimodal guidance using directional audio and distance-sensitive vibrations, allowing Brian to reach and retrieve his keys with confidence.

\textbf{Scenario 2: Requesting a Drink.}  
In another situation, Denise, a person with visual impairment, is feeling thirsty. Rather than navigating a cluttered app menu or switching between multiple tools, she simply says, “I’m thirsty” to NaviSense. The app responds, “Would you like me to find a water bottle or a soft drink?” Denise replies, “A water bottle is good,” prompting the system to scan her surroundings. Upon detecting the bottle, NaviSense provides continuous verbal and haptic cues to guide her hand directly to it, without typing or guesswork.

These examples demonstrate how NaviSense uniquely bridges the gap between open-world visual recognition and actionable, hands-on guidance. Unlike existing systems that either rely on pre-scanned objects or provide only scene-level descriptions, NaviSense supports real-time, autonomous object retrieval with no prior setup, adapting fluidly to dynamic and cluttered environments.

\section{Appendix: Expanded Formative Interview Insights}

To complement the main design considerations presented in the paper, we include additional user feedback gathered during the formative interview study with 12 participants (see Tab.~\ref{tab:participant-demo}) with visual impairments. These insights informed key design decisions in NaviSense and are organized around the three core design themes identified in our analysis.

\noindent\textbf{\textit{D1: Open-world Object Detection.}} Participants noted failures of existing systems in recognizing everyday items, particularly items with similar packaging. For instance, participants mentioned that current systems struggle to differentiate between similar items such as shampoo versus conditioner (P5). Further, Be My AI often fails to accurately describe colored objects (P8).

\noindent P9 talked about struggles with identifying medications:
\begin{quote}
    \textit{``With medicine bottles, especially the round ones, SeeingAI just doesn’t work. It either misses the label or gives me something completely wrong.''} (P9)
\end{quote}
\noindent P4 added that even common household items can be difficult for current systems:
\begin{quote}
    \textit{``Things like sunscreen or moisturizer tubes confuses Be My AI. And to identify food, say, brown rice or white rice, I have to try three different apps to figure out what it is.''} (P4)
\end{quote} 

\noindent\textbf{\textit{D2: Precise Guidance and Feedback.}}  Participants reported that while current systems might detect the presence of an object, they often fail to provide actionable navigational cues in real world scenarios. For example, P4 and P6 recounted difficulties with a microwave and elevator respectively: 

\begin{quote}
    \textit{``It was frustrating when I tried using Aira volunteer with a microwave; the volunteer had a hard time navigating me to the start or cancel button.''} (P4)
\end{quote}
\begin{quote}
    \textit{``I was on the elevator and couldn’t find the buttons to select the floor. Even after taking several pictures with Meta Glasses, I still couldn’t locate the buttons.''} (P6)
\end{quote}

\noindent Further, P11 described a scenario where, even though Be My AI detected a remote control on a desk, it provided no guidance to locate it amidst scattered objects:
\begin{quote}
    \textit{``I know the remote is on the desk, Be My AI says it is there on the table, but it doesn’t tell me where the remote is. It is no help in a scatter of objects.''} (P11)
\end{quote}

\textbf{\textit{D3: Conversational Interface.}}  Participants strongly preferred natural voice interaction over screen-based or typing interfaces. P4 emphasized the practicality of conversational approach, \textit{``Typing takes a lot of time; I’d rather just talk to an assistive device.''} P12 described this positively, stating, 
\begin{quote}
    \textit{``Conversational interaction with Meta glasess is like having a robot companion that you can talk to; this makes the whole interaction feel very natural.''} (P12)
\end{quote}

\section{Appendix: Implementation Details}

\begin{figure}[ht]
  \centering
  \includegraphics[width=0.9\linewidth]{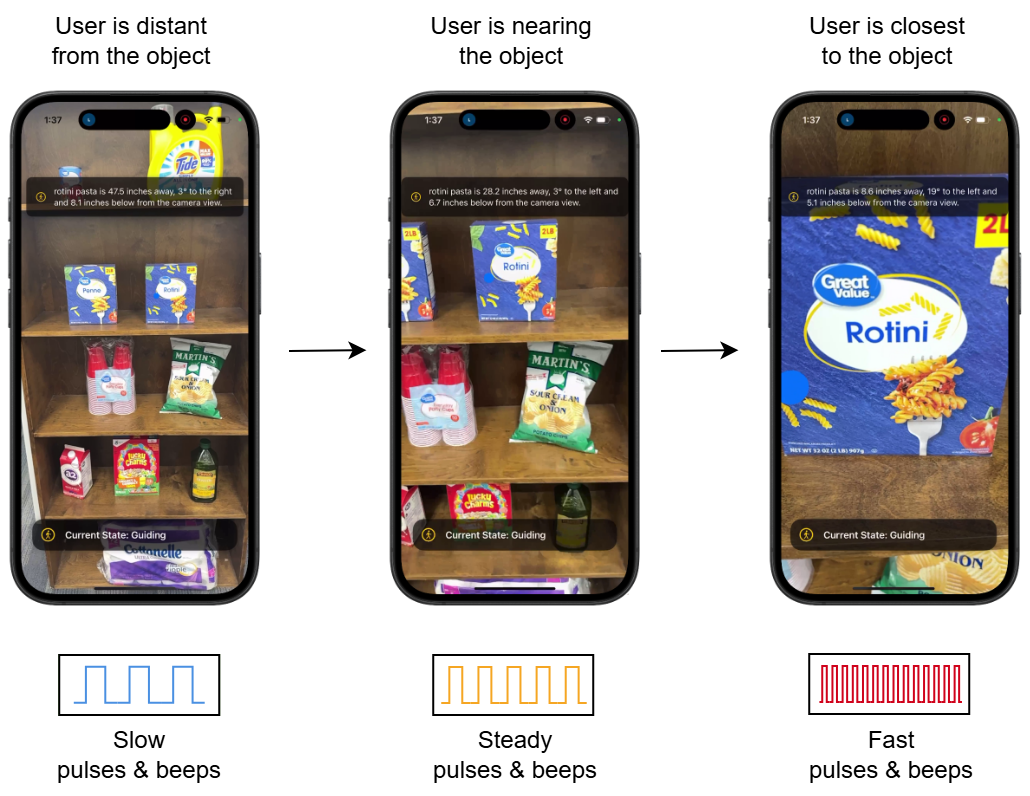}
  \caption{Multi-modal feedback in NaviSense during object guidance. The app provides real-time verbal instructions and uses haptic patterns (shown below each phone) to indicate distance to the target object. Left: user is far from the object (low-frequency haptics); Middle: getting closer (medium-frequency); Right: near the object (high-frequency).}
  \Description{Three smartphone screenshots show the NaviSense app guiding the user to a box of rotini pasta on a shelf. Each screen shows the distance, angle, and direction. Below each phone is a visual icon representing haptic pulse intensity—low, medium, and high—corresponding to the user's distance from the object.}
  \label{fig:multi-modal-feedback}
\end{figure}

This section elaborates on key system components of NaviSense. Figure~\ref{fig:multi-modal-feedback} illustrates the multimodal feedback patterns, i.e, slow pulses for distant objects, steady pulses when approaching, and rapid pulses upon reaching the target. Additionally, our cloud-based detection pipeline maintains low latency, averaging 0.706s per call with a P99 latency of 0.797s.

\begin{figure*}[t]
  \centering
  \includegraphics[width=\linewidth]{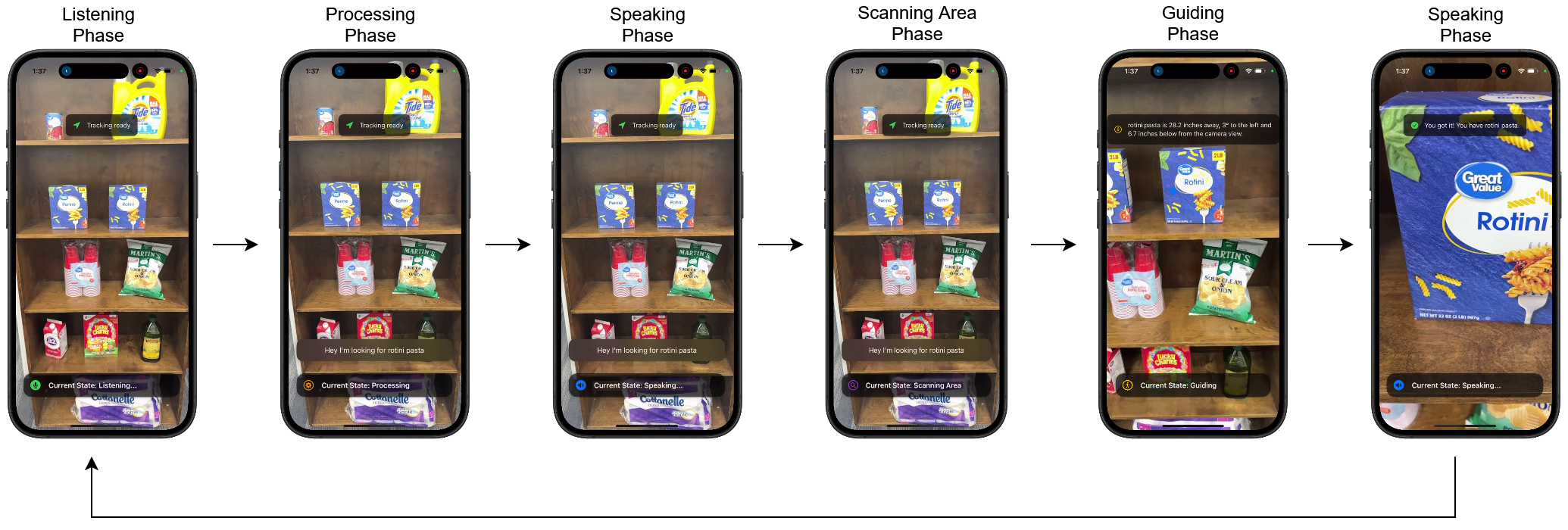}
  \caption{End-to-end object retrieval flow using NaviSense. The user initiates a search for rotini pasta using voice input and is guided in real time through various system states until successful retrieval is confirmed.}
  \Description{
  A horizontal sequence of six photographs of a smartphone, each showing a different stage of the NaviSense application during an object retrieval task. 
  In the first photo, the app is in "Listening" mode with a shelf of household items visible through the camera. 
  In the second, the user’s voice query “Hey, I’m looking for rotini pasta” is shown on-screen, and the system begins processing. 
  The third image shows the app speaking a response, while the fourth shows it scanning the environment for the item. 
  The fifth image captures the guidance phase, where spatial directions are overlaid to help the user find the rotini pasta. 
  The sixth and final photo displays a close-up of the rotini box with a success confirmation. 
  All phones are aligned in a left-to-right sequence with arrows indicating the interaction flow. The final image loops back to the first, illustrating that the process is repeatable.
  }
  \label{fig:usage_flow}
\end{figure*}

\subsection{Application State Model}

NaviSense follows a finite state machine architecture to manage user interaction and system behavior throughout the object retrieval process. This structured design supports modular development, improves system reliability, and provides a predictable, user-centered experience. A visual depiction of the state flow is presented in Fig.~\ref{fig:usage_flow}. Each application state is described below in sequence:

\begin{enumerate}
    \item \textbf{Idle}: The application initializes key system components, including ARKit, microphone access, and network connectivity. Once all services are ready, it transitions to the Listening state.
    
    \item \textbf{Listening}: The application actively monitors for voice commands. During this state, the user can speak naturally to specify the object they wish to find.
    
    \item \textbf{Processing}: After receiving voice input, the application transcribes the speech and sends the text to a backend server. A large language model (LLM) interprets the request and returns a natural language response that confirms the user’s intent.
    
    \item \textbf{Speaking}: The application conveys the LLM's response to the user through synthesized speech using the device’s built-in text-to-speech functionality.
    
    \item \textbf{Scanning}: The application enters a real-time scanning phase based on the response from processing state. It captures image frames and depth data at regular intervals (every 1 second in our case) to detect and localize the requested object using vision-language model inference.
    
    \item \textbf{Guiding}: Upon locating the object, the system initiates multi-modal guidance. It provides directional audio prompts and distance-sensitive haptic feedback to help the user navigate toward the object. Once the object is successfully reached, the system returns to the Listening state to await the next request.
\end{enumerate}

To support non-visual interaction, users can cancel or reset the current task at any time by shaking the device. This tactile gesture enables a quick return to the Listening state, eliminating the need for on-screen controls and ensuring accessible task management for users with visual impairments.

\section{Appendix: User Study Details}

\begin{table*}[ht]
\centering
\caption{Demographic information of participants with details about their visual impairment.}
\Description{This table presents demographic information for 13 participants. It includes details about each participant’s visual impairment, including the condition, onset, gender, age, and level of impairment.}
\label{tab:participant-demo}
\begin{tabular}{llllll}
\toprule
\textbf{PID} & \textbf{Gender} & \textbf{Age} & \textbf{Level of VI} & \textbf{VI Condition} & \textbf{Onset of VI} \\
\midrule
P1 & Female & 50 & Totally blind & Retinopathy of prematurity & Since birth \\
P2 & Male   & 62 & Totally blind & Aniridia & Since birth \\
P3 & Female & 67 & Totally blind & Retinitis Pigmentosa & Since birth \\
P4 & Female & 57 & Totally blind & Panuveitis & At age 26 \\
P5 & Female & 65 & Low vision & Diabetic retinopathy & Since birth \\
P6 & Female & 39 & Totally blind & Uveitis and retinal detachment & At age 6 \\
P7 & Female & 55 & Totally blind & Neuromyelitis optica & At age 36 \\
P8 & Female & 52 & Totally blind & Pseudotumor cerebri & At age 34 \\
P9 & Male & 73 & Low vision & Glaucoma & At age 39\\
P10 & Male & 55 & Totally blind & Retinitis Pigmentosa & At age 20 \\
P11 & Male & 46 & Low vision & Glaucoma & At age 31 \\
P12 & Female & 54 & Totally blind & Diabetic retinopathy & At age 42 \\
\bottomrule
\end{tabular}
\end{table*}

\subsection{Participants}

Table~\ref{tab:participant-demo} shows PVI participants' demographic information, including age, gender, visual impairment condition and age of onset. We recruited the participants through local chapters of the National Federation of the Blind using convenience and snowball sampling methods. Participants were eligible if they identified as blind or visually impaired and were familiar with the use of smartphones. 

\begin{figure}[t]
  \centering
  \includegraphics[width=0.4\linewidth]{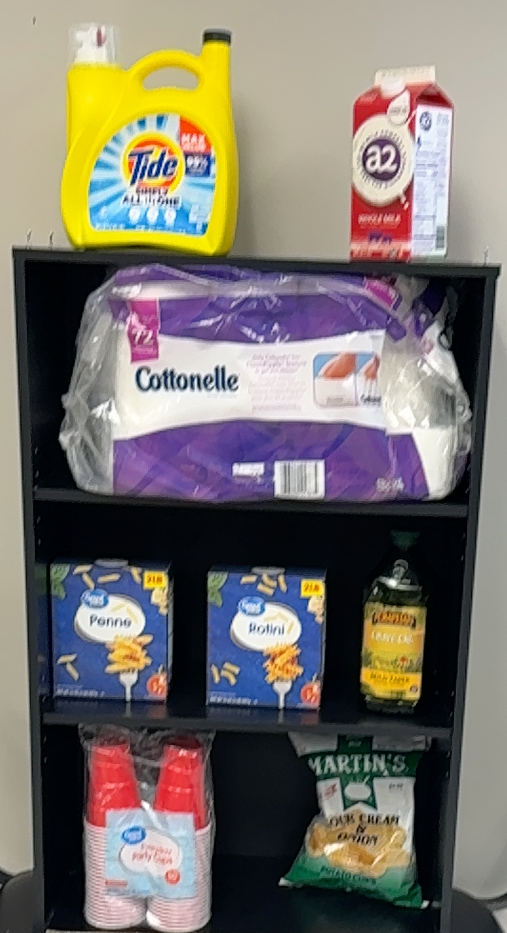}
  \caption{User Study Experiment Setup}
  \Description{A shelving unit used in the user study, containing various household objects. The top of the shelf holds a bottle of Tide laundry detergent and a carton of A2 milk. The second shelf has a large pack of Cottonelle toilet paper. The third shelf includes two boxes of pasta (penne and rotini) and a bottle of olive oil. The bottom shelf displays a stack of red plastic cups and a bag of sour cream and onion chips. Items are arranged in a cluttered yet visible configuration to simulate a realistic setting.}
  \label{fig:experiment_setup}
\end{figure}

\subsection{Detailed Study Procedure}

The user study's procedure and protocol was approved by our university's Institutional Review Board. The user study consisted of object retrieval tasks across three different system conditions:
\begin{itemize}
    \item \textbf{NaviSense}: Participants used our conversational AI system with audio-haptic guidance.
    \item \textbf{Be My AI}~\cite{bemyeyes_introducing_be_my_ai2023}: Participants used the Be My Eyes's AI feature (Be My AI) to take pictures of the scene, and based on the description given by the application, they tried to retrieve the object.
    \item \textbf{Ray-Ban Meta Smart Glasses}~\cite{raybanmeta}: The participants used the \textit{``Hey Meta! Look and Find \{object\}''} command to find an object, and based on the response, they retrieved the object.
\end{itemize}

\noindent We began the study by obtaining informed verbal consent from all participants. Before beginning the trials, each participant received a brief tutorial and orientation for every system. During each trial, participants stood approximately 5 feet from a wooden shelf consisting of four rows and stocked with eight commonly used household or grocery products (see Fig.~\ref{fig:experiment_setup}). The items included Martin’s Sour Cream \& Onion Potato Chips~\cite{martins_snacks}, Pompeian Olive Oil (Mild Taste)~\cite{pompeian_products}, Great Value Rotini Pasta (2 lb), Penne Pasta (2 lb), and Red Plastic Party Cups (50 count)~\cite{great_value_products}, Tide Laundry Detergent (All-In-One, HE)~\cite{tide}, an A2 Milk Carton~\cite{a2_milk}, and Cottonelle Toilet Paper (72 count)~\cite{cottonelle}.

Participants were asked to locate and retrieve three target items: a milk carton, red party cups, and rotini pasta. These were selected to vary in visual and tactile properties: the rotini pasta was visually similar to the penne pasta placed nearby, the red party cups were partially transparent and reflective, and the milk carton was distinct in both shape and size. Each participant completed the retrieval task using all three systems, with the order counterbalanced across participants to mitigate order effects. For each trial, participants first listened to the system’s prompt and then attempted to locate and retrieve the specified object. Object positions were randomized between trials, and each trial was independently timed and recorded.

After the testing session, participants were asked to complete a brief semi-structured interview covering system usability, ease of guidance, and their general experience.

\section{Appendix: Additional Results}

\subsection{Object-Specific Retrieval Times}
\begin{table}[ht]
\centering
\caption{Mean total task time (in seconds) with standard deviation, reported per method and object. Lower values indicate better performance.}
\Description{This table presents the total task completion time across three objects—A2 Milk, Party Cups, and Rotini Pasta—using three methods: BeMyEyes, MetaGlasses, and NaviSense. Each cell shows the average time taken in seconds followed by the standard deviation. NaviSense consistently shows the lowest values across all objects.}
\label{tab:accessible_total_time}
\begin{tabular}{lcccc}
\toprule
\textbf{Method} & \textbf{A2 Milk} & \textbf{Party Cups} & \textbf{Rotini Pasta} \\
\midrule
Be My Eyes        & 66.21 $\pm$ 25.95 & 58.76 $\pm$ 25.38 & 67.61 $\pm$ 23.44 \\
Meta Glasses     & 37.58 $\pm$ 17.65 & 39.70 $\pm$ 18.61 & 45.27 $\pm$ 25.43 \\
NaviSense & \textbf{30.94} $\mathbb{\pm}$ \textbf{8.07} & \textbf{31.56} $\mathbb{\pm}$ \textbf{8.64}  & \textbf{32.70} $\mathbb{\pm}$ \textbf{7.74}  \\
\bottomrule
\end{tabular}
\end{table}

Table~\ref{tab:accessible_total_time} presents the mean total task completion time, along with standard deviations, for each object retrieval task—A2 Milk, Party Cups, and Rotini Pasta—across the three assistive methods. Across all objects, NaviSense consistently demonstrates the fastest performance, which indicates that it not only supports efficient object identification but also enables users to complete retrieval tasks quickly and reliably. In addition to its speed, NaviSense also exhibited the lowest variability, as shown in its standard deviations. This consistency suggests a more reliable and predictable user experience across different participants and trials. In contrast, both Be My Eyes and Ray-Ban Meta Smart Glasses show higher variances, indicating more fluctuation in user performance—likely stemming from inconsistencies in verbal feedback, lack of adaptive support, or delays in system responsiveness. These results enhance the advantage of real-time adaptive guidance provided by NaviSense in supporting object recognition efficiency and reliability for blind and low-vision users.

\subsection{User Feedback}
We show additional user feedback that we received as part of the post study evaluation interview.
\begin{enumerate}
    \item Effectiveness of Audio and Haptic Feedback: 
    \begin{quote}
        \textit{``I like the fact that it is giving you cues to the location of the object is, whether it is left or right, up or down, and then bullseye, boom, you got it!''} (P3)
    \end{quote}
    \begin{quote}
    \textit{``The beeping sound, I love that. It's like that game you know, you're hot, you're cold, you're hot again. It really helps you zero in on object locations.''} (P7)
    \end{quote}
    
    \item Desire for Integrated Features and Customization:
    \begin{quote}
        \textit{``Honestly, I wish there was just one device that did it all. Like, the hands-free part of the Meta Glasses was great, but NaviSense had the best audio cues to actually guide me to the object. And maybe something like the short descriptions from Be My Eyes, but, you know, just more focused on where exactly the thing is.''} (P6)
    \end{quote}  
    \item Envisioned Future Use Cases and Enhancements:
    \begin{quote}
        \textit{``
    I like to cook, it could tell me, what I pull out of my freezer, with labels, steak vs liver in the freezer. Something to have help me from the airpot, finding my luggage in a conveyer belt. ''} (P8)
    \end{quote}
\end{enumerate}

\subsection{Evaluation on Everyday Object Categories}

\begin{figure}[t]
  \centering
  \includegraphics[width=0.40\linewidth]{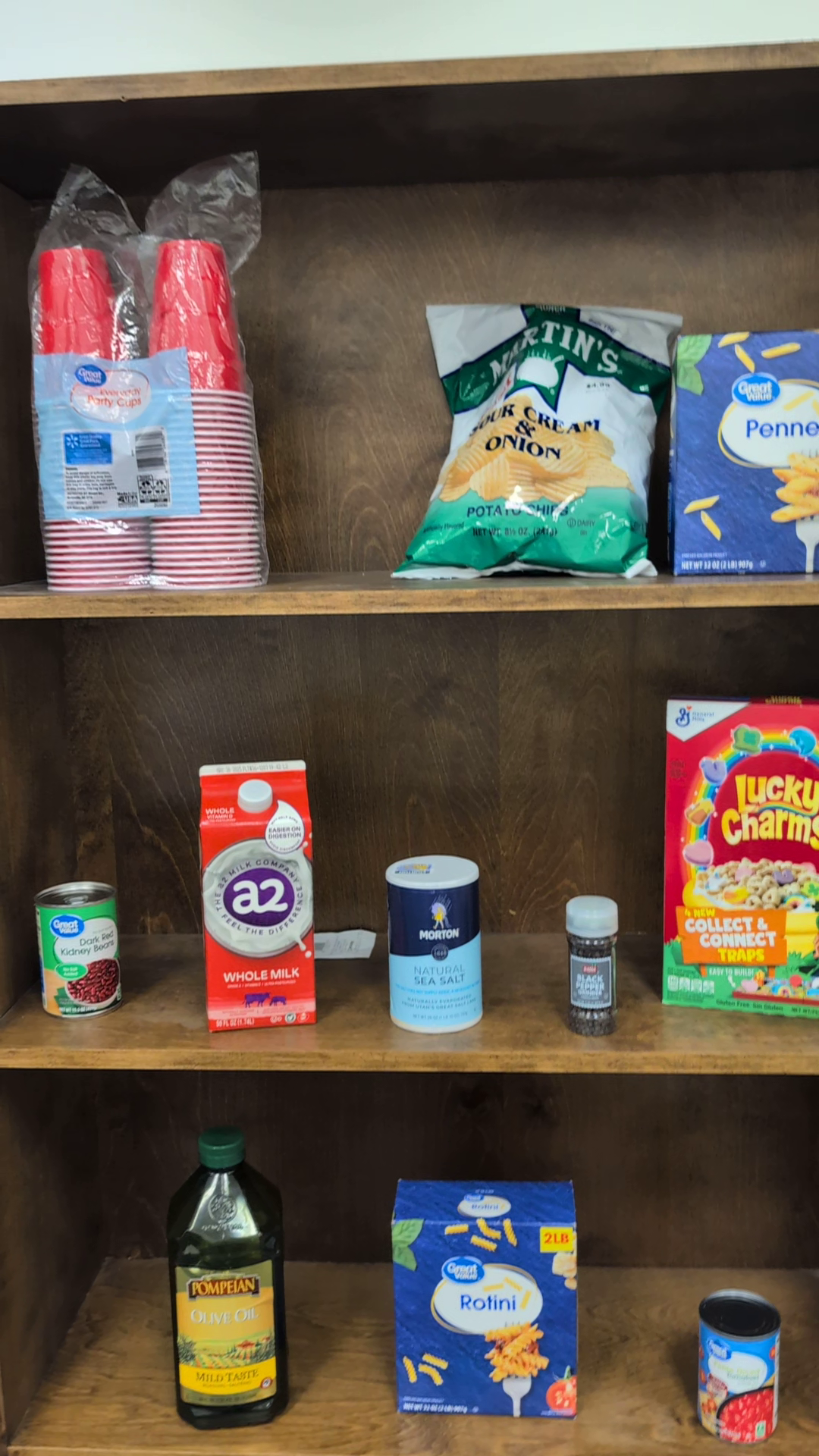}
  \includegraphics[width=0.40\linewidth]{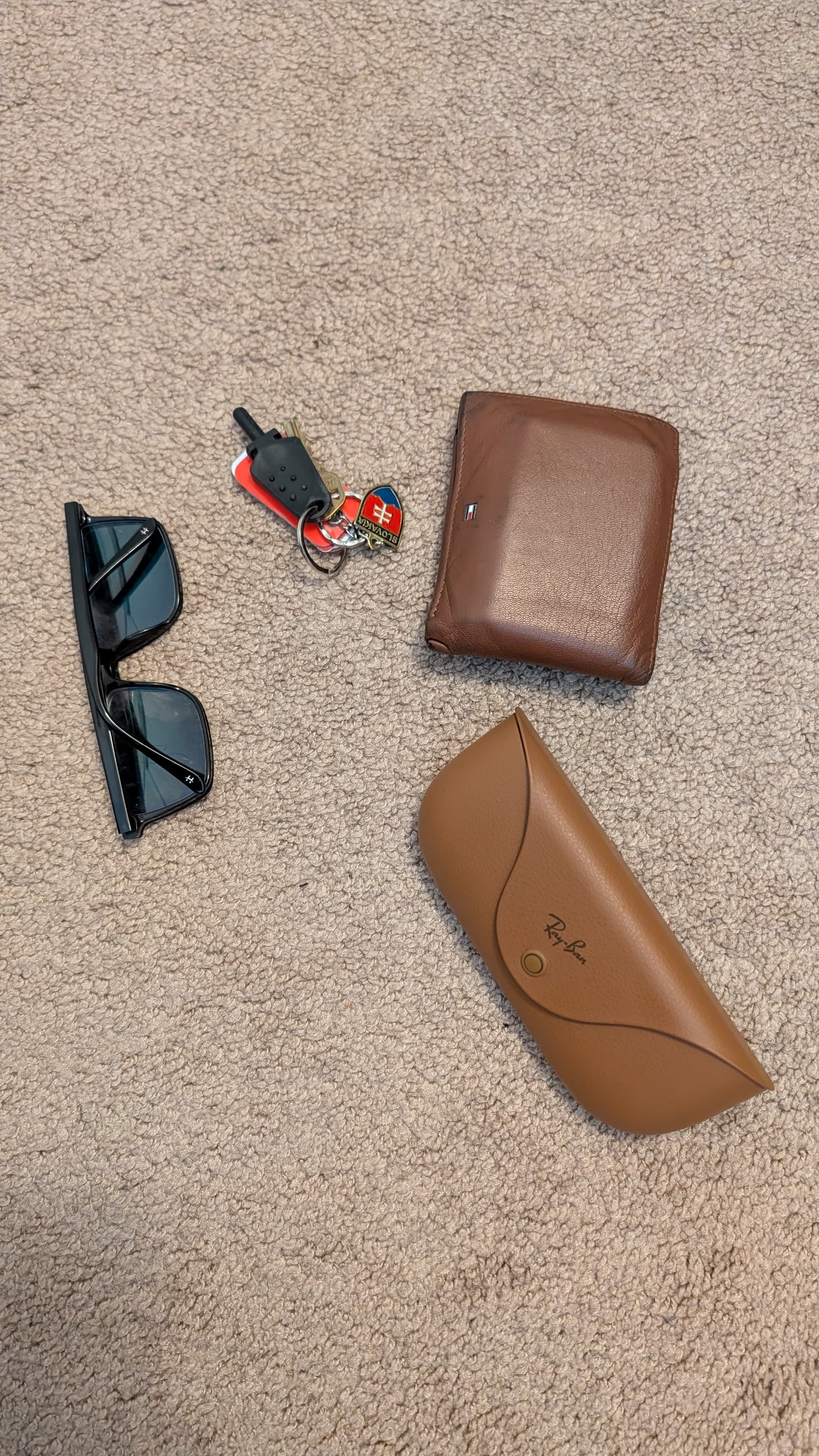}
  \caption{Sample images from the dataset where the objects are of different size and cluttered as well.}
  \Description{Two sample images from the NaviSense dataset. Left: a wooden shelf with various grocery items including stacked red plastic cups, bags of chips, pasta boxes, cereal, and canned goods arranged across three shelves. Right: a carpeted floor with scattered personal items such as a wallet, sunglasses, car keys, and a Ray-Ban glasses case. The scenes show variation in object type, size, and clutter.}
  \label{fig:ecological_validity}
\end{figure}

To complement the main evaluation, we include additional visual examples from our custom dataset in Fig.~\ref{fig:ecological_validity}. The dataset features scenes with varied object sizes, types, and degrees of clutter, including items not part of the original 25-object set used in our structured evaluation. This setup was designed to better reflect real-world household environments.

NaviSense demonstrated strong generalization under these conditions, achieving 95\% accuracy across 200 detection attempts (95\% Wilson CI = [0.911, 0.974]). The system accurately identified or ruled out the target object in 190 frames, with most errors resulting from occlusion or visual ambiguity. These findings reinforce the system’s robustness across a diverse range of household items and support its ecological validity. Future work will explore performance under broader environmental variations, including different lighting, object densities, and background complexity.

\section{Appendix: Technical and Design Iterations}

Several technical directions were explored during development but ultimately set aside due to constraints in cost, performance, or scalability. For instance, we initially prototyped a continuous voice interaction system using OpenAI’s Realtime API~\cite{openai_realtime_api}, which enabled more fluid and naturalistic exchanges. However, the approach proved prohibitively expensive for real-time, edge-based deployment. As a result, we adopted a more sustainable and modular pipeline—consisting of speech-to-text input, followed by an LLM-based reasoning layer, and concluding with text-to-speech output. This architecture provided greater control, cost-efficiency, and overall robustness in mobile settings.

We also evaluated several lightweight open-world object detection models, including GroundingDino~\cite{liu2024grounding}, DETR~\cite{zhu2020deformable}, and OWL-v2~\cite{minderer2023scaling}, selected for their potential to operate on-device. Despite their efficiency, these models struggled to recognize everyday household items with sufficient accuracy in early testing. To address this, we opted to use a cloud-based Vision-Language Model~\cite{zhang2024vision}, which enabled more flexible, open-ended, and user-defined object queries. While this choice significantly improved recognition performance, it also introduced trade-offs in latency and data privacy—highlighting key areas for future optimization.

\end{document}